\documentclass{icrc2009}

\usepackage{graphicx}   
\usepackage[caption=false]{caption}    
\usepackage[font=footnotesize]{subfig} 
\usepackage{fixltx2e}

\newcommand{\shorttitle}[1]%
{\markboth{Proceedings of the 31\MakeLowercase{$^{st}$} ICRC, {\L}\'{o}d\'{z} 2009}{#1} }
\newcommand{\etal}{\MakeLowercase{\textit{et al. }}} 

\usepackage{units}
\usepackage{amsmath}
\usepackage{url}
\usepackage{upgreek}

\begin{document}
\title{Event reconstruction with the proposed large area \\ Cherenkov air shower detector SCORE}

\author{\IEEEauthorblockN{Daniel Hampf\IEEEauthorrefmark{1},
			  Martin Tluczykont\IEEEauthorrefmark{1},
                          Dieter Horns\IEEEauthorrefmark{1}}
                            \\
\IEEEauthorblockA{\IEEEauthorrefmark{1}University of Hamburg, Physics Department, Luruper Chaussee 149, 22761 Hamburg}
}

\shorttitle{Daniel Hampf \etal: SCORE - Event Reconstruction}
\maketitle

\begin{abstract}
The proposed SCORE detector consists of a large array of light collecting modules designed to sample the Cherenkov light front of extensive air showers in order to detect high energy $\upgamma$-rays. A large spacing of the detector stations makes it possible to cover a huge area with a reasonable effort, thus achieving a good sensitivity up to energies of about a few 10 PeV.

In this paper the event reconstruction algorithm for SCORE is presented and used to obtain the anticipated performance of the detector in terms of angular resolution, energy resolution, shower depth resolution and $\upgamma$ / hadron separation.
\end{abstract}

 
\section{Introduction}
The SCORE detector is designed to study $\upgamma$-rays in the energy regime from a few $\unit[10]{TeV}$ up to some $\unit[10]{PeV}$. The detector consists of a large array of light sensitive detectors with a half opening angle of about $30^\circ$ and is used to sample the Cherenkov light shower front of extensive air showers (EAS). Each detector contains a few photomultiplier tubes (PMTs) which are equipped with Winston Cones, facing towards the sky. With a light sensitive area in the order of $\unit[1]{m^2}$ in the individual detector stations it will be possible to sample the shower front up to high distances from the shower core, thus making it possible to place the stations about $\unit[100]{m}$ apart from each other. The signals from the PMTs will be sampled by fast Flash-ADCs, making it possible to use not only the photon density but also timing features for event reconstruction.

A comprehensive overview of the physical motivation and the detector concept can be found in another contribution on this conference \cite{MT}. 

In this paper a scheme for event reconstruction and results of simulations testing this reconstruction algorithm are presented. 

\section{Simulations \label{simulations}}
The simulation chain consists of air shower simulations in COSIKSA \cite{Heck} (version 6.75), the detector simulation \itshape sim\_score \normalfont and a reconstruction framework. Showers induced by $\upgamma$-rays and protons in the energy range from $\unit[10]{TeV}$ up to $\unit[3]{PeV}$ have been simulated, and the results are analysed using different detector geometries. The two detector alternatives discussed here are:
\begin{enumerate}
 \item $\unit[100]{m}$ spacing and $\unit[0.5]{m^2}$ entrance window
 \item $\unit[200]{m}$ spacing and $\unit[1.5]{m^2}$ entrance window
\end{enumerate}
In each energy band showers with zenith angles of $0^\circ$, $15^\circ$ and $20^\circ$ have been simulated, and all results given here are averaged values over the different angles (the same number of showers has been simulated for each angle).

The detector simulation includes atmospheric absorption of Cherenkov light and the anticipated night sky background. The detector modules are modelled with a realistic PMT pulse shape function (taken from \cite{Henke}) and a wavelength dependent quantum efficiency (taken from \cite{datasheet_ET}). The angular acceptance of the Winston Cones has been studied with ray tracing simulations and incorporated into the detector simulation.

All resolutions given in here are $1\upsigma$ values of the difference between MC input and the reconstructed quantity.

\section{Measured quantities \label{data}}
\begin{figure}[!t]
  \centering
  \includegraphics[width=2.8in]{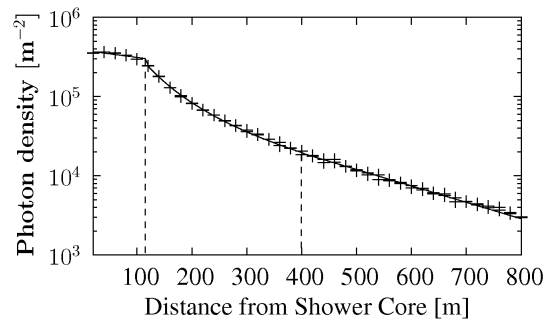}
  \caption{\textbf{Photon density distribution} with the described fit function for an EAS induced by a $\unit[1]{PeV}$ $\upgamma$-particle. For demonstration purposes a $\unit[20]{m}$ sampling is assumed.}
  \label{lat_dist}
 \end{figure}

Firstly, the lateral distribution of the photon density is recorded. It can be seen that the intensity falls off exponentially up to about $\unit[120]{m}$ from the shower core. From $\unit[120]{m}$ to $\unit[400]{m}$ the light distribution follows a power law, until it changes to an exponential function again (see Fig. \ref{lat_dist}):
\begin{equation}
   f(r)=
   \begin{cases}
      	P_1 \exp (d_1 r) & \mbox{for } r < \unit[120]{m} \\
	Q (\frac{r}{\unit[220]{m}})^k & \mbox{for } \unit[120]{m} < r < \unit[400]{m} \\ 
	P_2 \exp(d_2 r) & \mbox{for } r > \unit[400]{m} \\ 
   \end{cases}     
   \label{LDF}
\end{equation}

The part up to $\unit[120]{m}$ can only be used if a spacing of $\unit[100]{m}$ is assumed, otherwise there are not enough data points in this range. The data points beyond $\unit[400]{m}$ usually have a quite poor signal to noise ratio, making the power law part the most important for SCORE. 

If continuity at the connection points ($\unit[120]{m}$ and $\unit[400]{m}$) is required in the fits, the number of free parameters can be reduced to four: $Q$, $k$, $d_1$ and $d_2$. Typical values for the power law index $k$ are in the range of -2 to -2.6.

\begin{figure}[!t]
  \centering
  \includegraphics[width=2.8in]{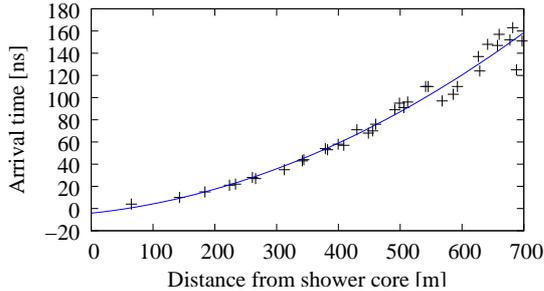}
  \caption{\textbf{Arrival time distribution} with parabolic fit function. Shown is the simulation of an EAS of a $\unit[2]{PeV}$ $\upgamma$-particle, recorded by a detector array with a spacing of $\unit[200]{m}$.}
  \label{peaks}
 \end{figure}

\begin{figure}[!t]
  \centering
  \includegraphics[width=2.8in]{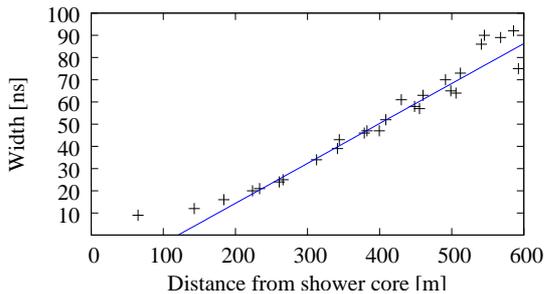}
  \caption{\textbf{Width distribution} for the same shower as in Fig. \ref{peaks} with a linear fit from $\unit[200]{m}$ to $\unit[500]{m}$.}
  \label{widths}
 \end{figure}

Secondly, the peak (highest entry in signal histogram) and the width of each signal (FWHM) are determined.  
Plotting the peak times versus the distance from the shower core one finds that a parabolic function describes the data very well (Fig. \ref{peaks}). The width of the timing signal increases linearly at distances greater than about $\unit[200]{m}$, while at lower distances the width is roughly constant due to the time resolution of the PMTs (Fig. \ref{widths}).

\section{Shower Core Reconstruction \label{core_reco}}
The simplest approach for the reconstruction of the shower core is to calculate the centre of gravity of the measured light intensities (used for example in AIROBICC, see \cite{Karle}). While being robust and fast, this method can achieve only mediocre resolutions as it does not take into account the non-linearity of the lateral photon density function (LDF) shown in Fig. \ref{lat_dist}. 

\begin{figure}[!t]
  \centering
  \includegraphics[width=2.8in]{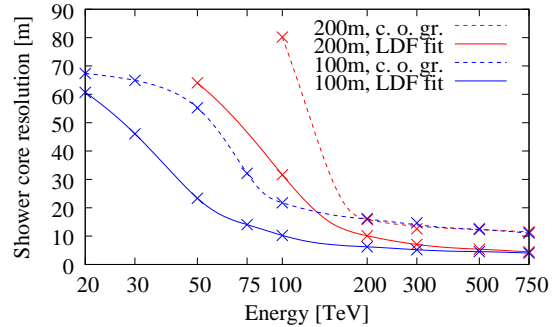}
  \caption{\textbf{Resolution of shower core} reconstruction versus $\upgamma$-ray energy for two different detector layouts with $\unit[100]{m}$ (blue) and $\unit[200]{m}$ (red) spacing (see section \ref{simulations}) and two reconstruction methods: centre of gravity method (dashed lines) and LDF fit (solid lines).}
  \label{core_resolution}
 \end{figure}

A better resolution can be achieved if the known LDF is fitted two-dimensionally to the recorded light intensities. Using the power law part of the LDF (see eq. \ref{LDF}), a resolution of better than $\unit[5]{m}$ can be achieved at higher energies. Towards lower energies both methods become less accurate when the signal approaches the noise level. While at higher energies the spacing of the detector stations makes no difference, the smaller spacing significantly decreases the threshold for a good reconstruction (see Fig. \ref{core_resolution}).

\section{Energy of Primary Particle}
\begin{figure}[!t]
  \centering
  \includegraphics[width=2.8in]{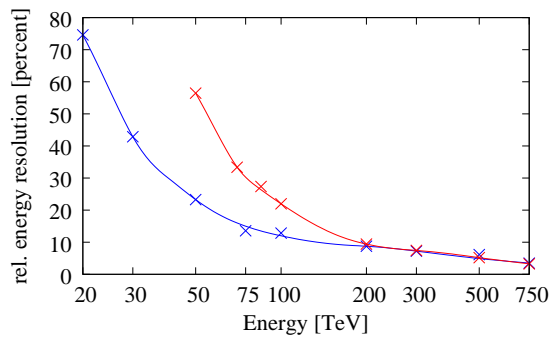}
  \caption{\textbf{Energy resolution} versus $\upgamma$-ray energy for two different detector layouts with $\unit[100]{m}$ (blue) and $\unit[200]{m}$ (red) spacing of stations and different sensitive areas per station (see section \ref{simulations}).}
  \label{energy_resolution}
\end{figure}

Simulations show that the photon density at any point of the Cherenkov shower front is proportional to the energy of the primary particle, which makes the energy reconstruction a straightforward process. However, at small distances from the shower core the photon density is also affected by the depth of the shower, while at large distances the signal to noise ratio gets worse. The best trade-off has been found at a distance of $\unit[220]{m}$, independent of the detector geometry. Taking the functional value at $\unit[220]{m}$ of the fitted LDF (eq. \ref{LDF}), an energy resolution of better than 10\% can be achieved at high energies. At lower energies, the detector variant with $\unit[100]{m}$ spacing performs significantly better due to the four times larger number of data points in the same distance interval, providing a better signal to noise ratio (see Fig. \ref{energy_resolution}). 

\begin{figure}[!t]
  \centering
  \includegraphics[width=2.8in]{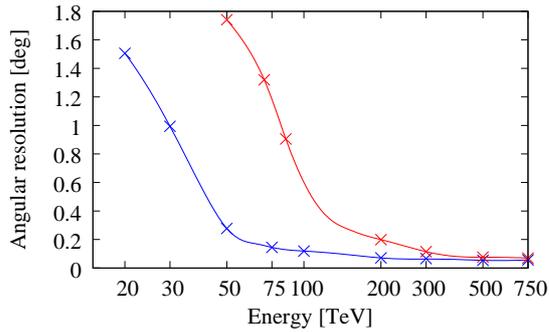}
  \caption{\textbf{Angular resolution} versus $\upgamma$-ray energy for two different detector layouts with $\unit[100]{m}$ (blue) and $\unit[200]{m}$ (red) spacing of stations and different sensitive areas per station (see section \ref{simulations}).}
  \label{wa}
\end{figure}

\section{Direction of Primary Particle}
It can be seen from Fig. \ref{peaks} that the peak arrival times are distributed according to a parabolic function for a vertical shower. For non-vertical showers, this parabola is tilted by the angle of the incoming particle. For finding the direction of the particle, the superposition of a plane and a three-dimensional parabola is fitted to the data points:
\begin{equation}
  f(x,y) = E_x x + E_y y + A (x^2 + y^2) + c\\
\end{equation}
$E_x$ and $E_y$ can then be used to calculate zenith and azimuth angle.

Both detector layouts achieve an angular resolution of better than $\unit[0.1]{^\circ}$ at higher energies, while the detector with the smaller spacing has a better performance at lower energies again (Fig. \ref{wa}).

\section{Shower Depth \label{section_depth}}
Several methods for the calculation of the shower depth are discussed in the literature. The TUNKA collaboration has shown that both the steepness of the inner part of the LDF and the widths of the timing signals (see sec. \ref{data}) can be used for this purpose \cite{TUNKA}. These methods are called intensity method and width method from here on. Simulations for SCORE show that also the time delay of the peak arrival times, which is fitted by a parabola (Fig. \ref{peaks}) can be used, if an accurate inter-station time calibration can be implemented (called timing method). Combining the different methods improves the depth resolution significantly.

In order to improve the performance of both the width and the timing method, all signals within a certain distance interval from the shower core can be summed up before determining peak position and signal width. As the number of detector stations available in a certain distance interval is proportional to the distance to the shower core, this procedure makes it possible to sample the width and the peak times up to high distances from the shower core. The size of the distance intervals is a critical parameter of the reconstruction and has to be optimized for each detector layout and energy range.

If the signals are still too noisy for a good determination of the width, a lognormal function can be fitted to the signals (see \cite{Chitnis}). Especially the determination of the width of the signal is improved significantly by that. 

\begin{figure}[!t]
  \centering
  \includegraphics[width=2.8in]{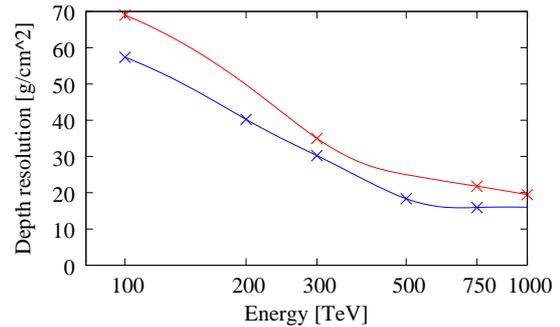}
  \caption{\textbf{Shower depth resolution} versus $\upgamma$-ray energy for two different detector layouts with $\unit[100]{m}$ (blue) and $\unit[200]{m}$ (red) spacing of stations and different sensitive areas per station (see section \ref{simulations}).}
  \label{depth_resolution}
\end{figure}

Figure \ref{depth_resolution} shows the depth resolution for the two different detector layouts. The $\unit[100]{m}$ detector variant uses all three described methods, while the $\unit[200]{m}$ detector can use only the width and the timing method, as there are not enough intensity data points at the inner part of the LDF. Both detectors achieve a resolution of better than about $\unit[30]{g / cm^2}$ at energies above $\unit[300]{TeV}$, but the detector with the smaller spacing performs better at all energies as it can combine all three methods for determining the shower depth.

\section{$\upgamma$ / hadron Separation}
The power to distinguish air showers of $\upgamma$-photons and hadrons is an important contribution to the sensitivity of a $\upgamma$-ray detector. The usual measure to quantify this power is the quality factor, which is defined as
\begin{equation}
  \mbox{QF} = \frac{\epsilon_{\upgamma}}{\sqrt{\epsilon_{hadron}}}
\end{equation}
where $\epsilon_p$ is the fraction of particles of the type $p$ which survive the applied cuts. Obviously, it is desirable to have a high chance to identify a $\upgamma$-photon as signal ($\epsilon_{\upgamma}$ close to one) and a low chance to falsely identify a hadron as signal ($\epsilon_{hadron}$ as small as possible), thus a high quality factor. 

Two approaches for $\upgamma$ / hadron separation are investigated:

\subsection{Longitudinal development}
It has been noted in section \ref{section_depth} that the shower depth can be obtained in different ways. While they all work well with both $\upgamma$ and hadron showers, it seems that the timing method and the width method produce slightly different results for different particles. This is believed to be due to the differences in the longitudinal development of showers of photonic or hadronic origin. While photonic showers tend to emit most light near the shower maximum, the light from hadronic showers is emitted from a wider range of altitudes. This makes the signals of hadronic showers slightly longer and leads to an overestimation of the shower depth for hadronic showers in the widths method.

\begin{figure}[!t]
  \centering
  \includegraphics[width=2.8in]{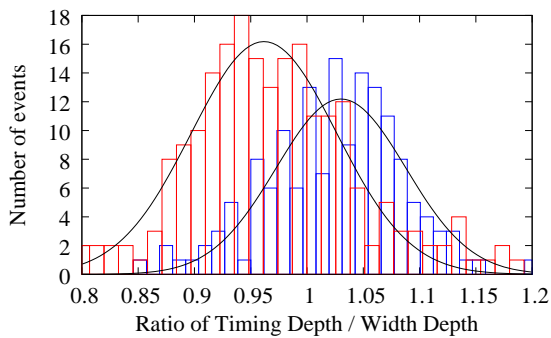}
  \caption{Ratio of the depth values obtained by timing method and width method for protons (red) and $\upgamma$-rays (blue) with energies between $\unit[500]{TeV}$ and $\unit[750]{TeV}$. The quality factor for a separation at 1 is about 1.2.}
  \label{gamma_hadron_manuell}
\end{figure}

 Comparing the two depth values can therefore be used as an indication for the nature of the particle. At higher energies (where both methods have good resolutions) this can be used for a basic $\upgamma$ / hadron separation. Figure \ref{gamma_hadron_manuell} shows histograms of the ratio of the depth values obtained by the two methods for $\upgamma$ and hadron showers. A quality factor of about 1.2 can be reached for energies above $\unit[500]{TeV}$.

\subsection{Multivariate Analysis}
In order to find a structure in the data that may improve the power for $\upgamma$ / hadron separation, self-learning algorithms can be used. So far, a boosted decision tree algorithm with $\upepsilon$-boosting (for reference, see e.g. \cite{decisiontree}) has been implemented and first tests look promising. Further tests with different input variables, algorithm parameters and algorithms are under way.

\section{Conclusion and Outlook}
The simulations conducted so far have shown that a shower front detector like SCORE can achieve angular and energy resolutions of the same level as current imaging Cherenkov telescopes (better than $0.1^\circ$ and $10\%$ respectively). Also, a satisfactory shower depth resolution of better than $\unit[30]{g/cm^2}$ has been found. Currently, the quality factor for background rejection is only at about 1.2, but this has to be considered as a first preliminary result. 
Optimizations of the shower depth calculation algorithm are in progress (e.g. better fit routines and functions, cuts on bad data points), and it is anticipated that a better shower depth resolution will have a positive impact on the $\upgamma$ / hadron separation. 
Furthermore, other analytic and self-learning algorithms for background rejection are being explored currently.
\newpage

Of the two detector layouts presented here the one with the smaller spacing and the smaller sensitive area per detector station has been found to be superior in terms of reconstruction power at lower energies. However, it uses about 25\% more PMTs and four times more detector stations. Therefore, a reasonable approach might be to combine the two layouts, i.e. use a part of the detector section with smaller spacing for detection of lower energy particles and employ a high energy extension for sensitivity at highest energies. Currently, further simulations with different detector layouts and spacings are under way.

The aim of this part of the project is to find the optimal detector configuration for the desired application and calculate a sensitivity curve based on the simulated reconstruction power of this instrument.

\section{Acknowledgement}
Daniel Hampf acknowledges the support of the German Ministry for Education and Research (contract number 05A08GU1).

\end{document}